\documentclass[aps,pra,groupaddress,twocolumn,floatfix]{revtex4-2}
\usepackage{units}
\usepackage{amsmath}
\usepackage{amssymb}
\usepackage{graphicx}
\usepackage{bm}
\usepackage{multirow,color,relsize,ulem,microtype}

\newcommand{\be}{\begin{equation}}
\newcommand{\ee}{\end{equation}}

\newcommand{\re}[1]{\text{Re}[#1]}
\newcommand{\im}[1]{\text{Im}[#1]}

\newcommand{\cc}[1]{{\color{black}#1}}

\makeatletter
\renewcommand*\env@matrix[1][*\c@MaxMatrixCols c]{%
  \hskip -\arraycolsep
  \let\@ifnextchar\new@ifnextchar
  \array{#1}}
\makeatother

\begin{document}

\title{Effective size of a parity-time symmetric dimer}

\author{Li Ge}
%\email{li.ge@csi.cuny.edu}
\affiliation{\textls[-18]{Department of Physics and Astronomy, College of Staten Island, CUNY, Staten Island, NY 10314, USA}}
\affiliation{The Graduate Center, CUNY, New York, NY 10016, USA}

\date{\today}

\begin{abstract}
Parity-time (PT) symmetric dimers were introduced to highlight the unusual properties of non-Hermitian systems that are invariant after a combined parity and time reversal operation. They are also the building blocks of a variety of symmetry and topologically protected structures, especially on integrated photonic platforms. As the name suggests, it consists of two coupled oscillators, which can be optical, mechanical, electronic, and so on in nature. In this article, we show that its effective size, defined by the number of lattice sites inversely proportional to the lattice momentum, is surprisingly three instead of two from the perspective of energy quantization. More specifically, we show analytically that the complex energy levels of a one-dimensional concatenated chain with $N$ PT-dimers are determined by a system size of $1+2N$, which reduces to three in the case of a single PT-dimer. We note that while energy quantization conditions were established in various non-Hermitian systems, exact and explicitly quantized complex energies as reported here are still scarce. In connection, we also discuss the other symmetries of a PT-dimer and concatenated PT-dimer chain, including non-Hermitian particle-hole symmetry and chiral symmetry. 
\end{abstract}

\maketitle

\section{Introduction}

\begin{figure*}
\includegraphics[clip,width=\linewidth]{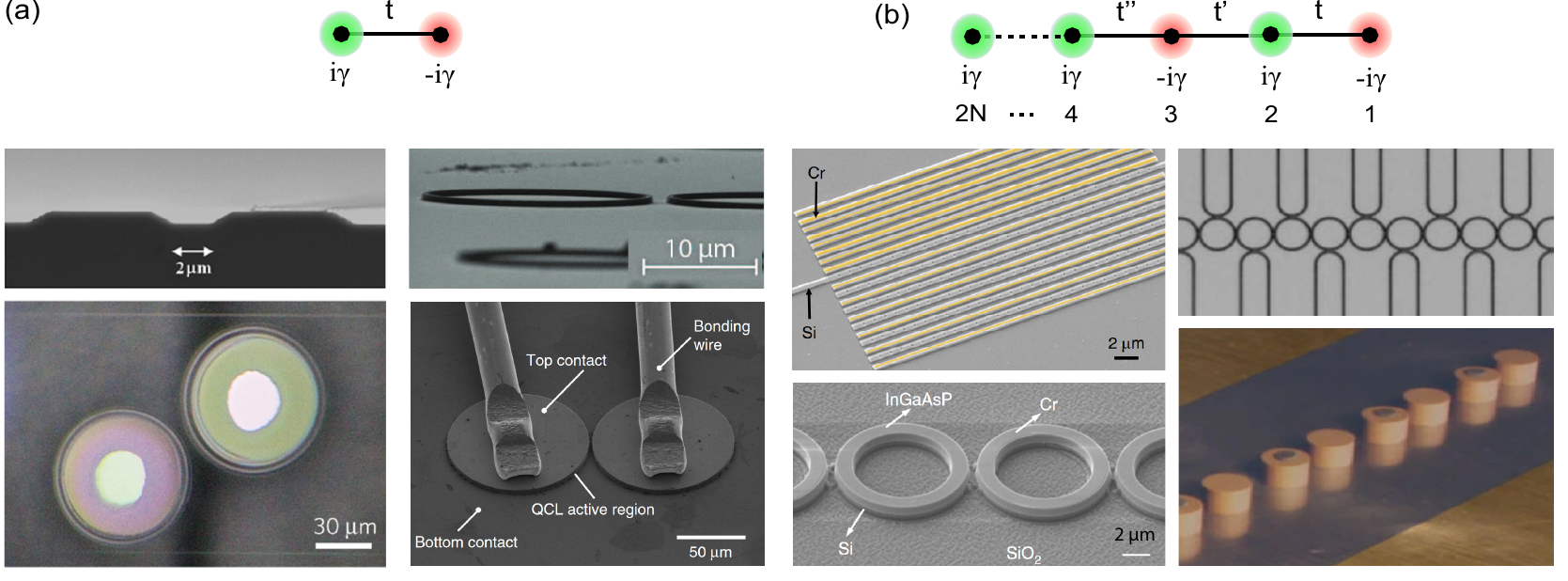}
\caption{Schematics and representative integrated photonic realizations of (a) a PT-dimer and (b) a concatenation of PT-dimers. The experimental figures are adapted from Refs.~\cite{Guo,Hodaei,Peng,Brandstetter,Poli,Parto,Zhao,Pan} with permissions. We take all couplings in (b) to be the same in the main text.} \label{fig:schematic}
\end{figure*}

A parity-time (PT) symmetric dimer is one of the most well known non-Hermitian model exhibiting PT symmetry \cite{Bender1,Bender2}, which is invariant after a combined parity and time reversal operation. It is typically described using a two-by-two effective Hamiltonian $H$, with identical real coupling coefficients between two oscillators and opposite imaginary on-site potentials, representing the rates of energy input and output. A PT-dimer was first realized in integrated photonics (see Fig.~\ref{fig:schematic}a with Refs.~\cite{Guo,Hodaei,Peng,Brandstetter}), using evanescently waveguides and microcavities on AlGaAs, LiNbO$_3$, and erbium-doped silica platforms \cite{NPReview}. Subsequently, it was also realized in mechanical \cite{Bender_Exp}, plasmonic \cite{Lawrence}, acoustic \cite{acoustic}, and electronic systems \cite{electronic}.  
The intriguing properties of a PT-dimer include the existence of exceptional points (EPs) \cite{EPreview}, which are non-Hermitian degeneracies with unequal geometric and algebraic multiplicities \cite{reveal}. Across its EPs, the two energy levels of a PT-dimer change from real to be complex conjugates, known as PT-symmetric and PT-broken phases, respectively.    

PT-dimers are also the building blocks of various symmetry and topologically protected structures, especially on integrated photonic platforms. For example, a layer of side-by-side PT-dimers in a two-dimensional photonic insulator laser defines the effective boundary of the system and can be used to route chiral edge states on demand \cite{steering}. A one-dimensional (1D) chain of PT-dimers with alternate orientations (e.g., gain-to-gain between the first two dimers and loss-to-loss between the second and third dimers) creates dimerization and also topological edge state(s) \cite{ptchain}, just as in a Hermitian Su-Schrieffer-Heeger (SSH) chain. Similar 1D structures with photonic edge or interface modes can also be constructed without flipping the orientation of the PT-dimers \cite{Poli,Parto,Zhao,Pan} (see Fig.~\ref{fig:schematic}b). 

In this article, we focus on a different and seemingly apparent question: What is the effective size of a PT-symmetric dimer, \cc{defined by the number of lattice sites inversely proportional to the lattice momentum? This question is important in understanding the discretized energy levels of a finite-sized system, whether it is Hermitian or non-Hermitian.} Given that a PT-dimer consists of two coupled oscillators, one may immediately jump to the conclusion that the answer is two. However, as we will show below, its appropriate size is three instead of two from the perspective of energy quantization. This statement, of course, requires clarification, as one does not observe a factor of three (or even two) in the two energy eigenvalues of a PT-dimer. However, once we consider a concatenation of $N$ PT-dimers with the same coupling, a crucial factor given by $1+2N$ emerges that represents the size of the chain, which becomes three in the case of a single PT-dimer. 

We note that while energy quantization conditions for non-Hermitian systems have been established in various scenarios, including high-precision approximations \cite{Bender2}, an imaginary coordinate shift of exactly solvable Hermitian potentials \cite{Gezai}, and the Milne quantization and supersymmetry algebra \cite{Dey}, exact and explicitly quantized complex energies as reported here are still scarce. In fact, one may have wondered whether such quantized complex energies, determined by an integer quantum number, exist at all in a simple non-Hermitian system, and our finding provides an affirmative answer to this question.
%Despite its small size, we also mention that a PT-symmetric dimer has other symmetries as well, 
\cc{We also note that although most PT-dimers and chains demonstrated so far experimentally are classical (see, e.g., Fig.~\ref{fig:schematic} and Ref.~\cite{NPReview}), the energy quantization procedure is the same as in non-interacting quantum systems due to their shared wave nature. For example, the evanescent coupling between two optical elements plays the same role as the tunneling of electrons between two atomic orbitals, and hence the respective energy bands they induce can be treated on the same footing.} In connection, we also discuss the other symmetries of a PT-dimer and the aforementioned PT-dimer chain, 
including non-Hermitian particle-hole symmetry \cite{zeromodeLaser,Malzard,NHFlatband_PR,NHFlatband_PRL,noether} and chiral symmetry \cite{chiral}.    

\section{Model}

Using $\pm i\gamma$ and \cc{$t$ (as in ``tunneling'' for evanescently coupled optical elements)} to denote the imaginary potentials (i.e., gain and loss) and real coupling, the effective Hamiltonian of a PT-dimer can be written as
\be
H = 
\begin{pmatrix}
i\gamma & \cc{t} \\
\cc{t} & -i\gamma
\end{pmatrix}\quad(t,\gamma>0),
\ee
and its two energy levels are given by
\be
E_\pm = \pm\sqrt{\cc{t}^2-\gamma^2}.\label{eq:E_dimer}
\ee
If we are just given the expression for $|E_\pm|$, then there is no indicator for the size of the system. Therefore, to define the size of a PT-dimer, we resort to the standard reference used in elementary quantum mechanics, which states that the energy levels inside a 1D box potential depend on the inverse length of the box. 

\begin{figure*}
\includegraphics[clip,width=0.8\linewidth]{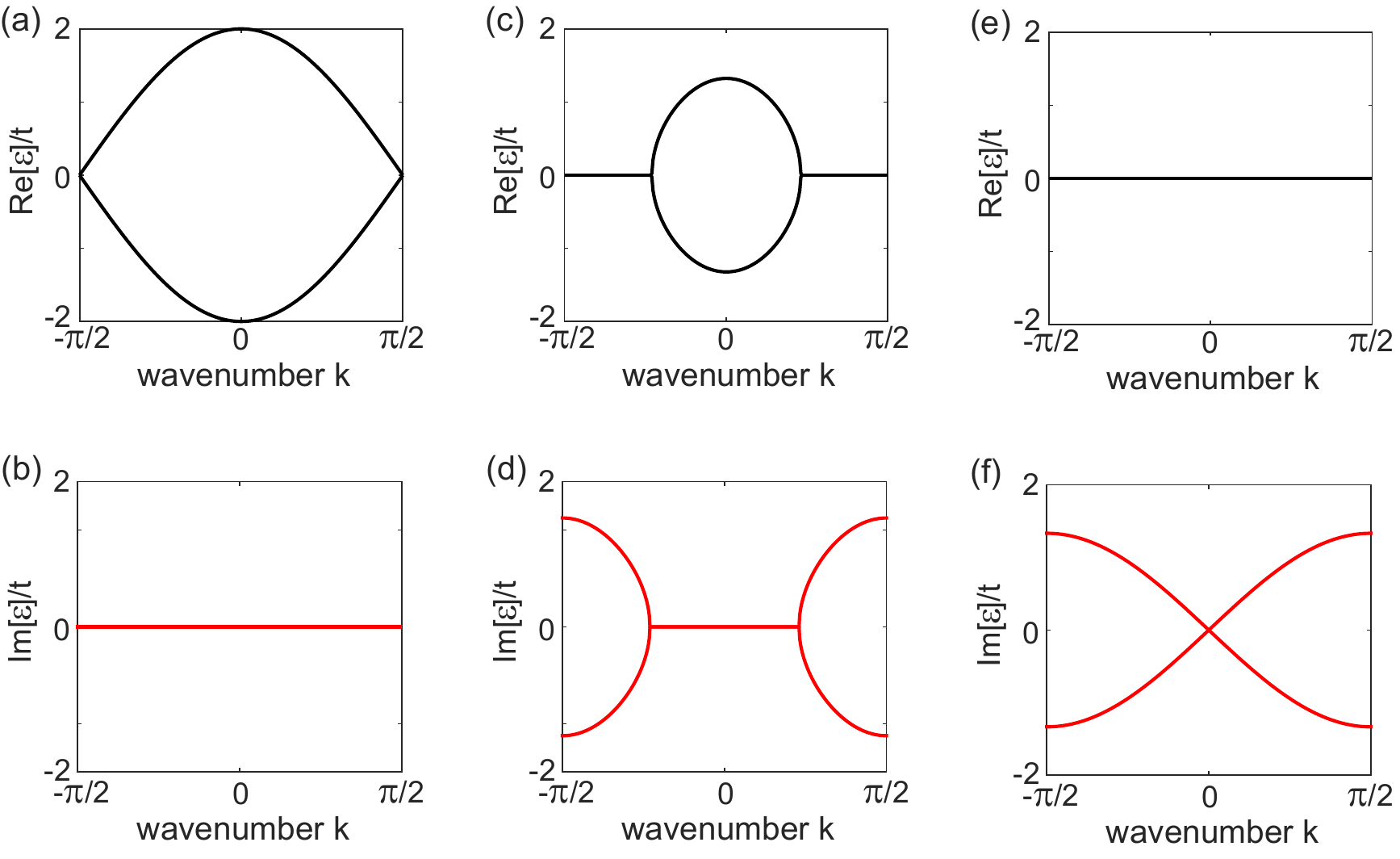}
\caption{Real and imaginary parts of the band structure of a concatenated PT-dimer chain with periodic boundary condition. The spacing between the two sites in a PT-dimer is taken to be 1, and $\gamma/t=0$ in (a,b), 1.5 in (c,d), and 2 in (e,f).} \label{fig:E_periodic}
\end{figure*}

To construct the analogy of a 1D box potential applicable to a PT-dimer, we consider the concatenation of PT-dimers as shown in Fig.~\ref{fig:schematic}b \cite{NHFlatband_PRL,Klaiman,linear}. With a unit cell of two sites and the periodic boundary condition, \cc{its Hamiltonian is given by
\be
H = 
\begin{bmatrix}
i\gamma & t(1+e^{-2ik}) \\
t(1+e^{2ik}) & -i\gamma
\end{bmatrix},
\ee
where $k\in[-\pi,\pi]$ is the lattice momentum along the chain and the spacing between two sites is taken to be 1 and dimensionless}. The two energy bands of the PT-dimer chain can be easily found to be 
\be
\varepsilon_\pm(k)=\pm i\sqrt{\gamma^2-2t^2(1+\cos k)},\label{eq:E_periodic}
\ee
which are either real or imaginary, with a maximum $|\re{\varepsilon_\pm}| = \sqrt{4t^2-\gamma^2}$ and a maximum $|\im{\varepsilon_\pm}| = \gamma$ when $\gamma\in[0,2t]$. For example, when $\gamma=0$, the chain is Hermitian and its unit cell contains a single site. Therefore, the two bands given by Eq.~(\ref{fig:schematic}) are real [see Figs.~\ref{fig:E_periodic}(a,b)], and they are formed by artificially folding the single energy band of the system at the middle of its first Brillouin zone. As $\gamma$ increases, the real parts of the two bands become the same near the edge of the first Brillouin zone \cite{makris}, where their imaginary parts become opposite [see Figs.~\ref{fig:E_periodic}(c,d)]. When $\gamma\geq2t$, the two bands become entirely imaginary, while their real parts become flat and identical [see Figs.~\ref{fig:E_periodic}(e,f)] \cite{NHFlatband_PRL}. 

As we will discuss later, this distribution of complex energy levels are due to multiple symmetries \cite{NHFlatband_PRL}. For now, we comment that for a PT-dimer chain of a finite size, if we can find its discrete energy levels analytically and if they have a similar form to Eq.~(\ref{eq:E_periodic}), then the argument of the cosine function in this analytical expression should be inversely proportional to the (effective) size of the PT-dimer chain.

To derive this analytical expression for the complex energy levels, we consider $N$ identical PT-dimers in the chain and label the lattice sites by $n=1$, 2, $\ldots$, $2N$ from right to left. \cc{The Hamiltonian of this finite-sized system is given by
\be
H = \sum_{n} (-)^ni\gamma|n\rangle\langle n| \,+\, t|n+1\rangle\langle n| + t|n\rangle\langle n+1|\label{eq:TBM}
\ee
in the second quantized form using the position basis.} 

\section{Energy quantization}
\label{sec:quantization}
For an eigenstate of $H$ given by Eq.~(\ref{eq:TBM}), e.g., 
\be
H\bm{\Psi}_\mu = \varepsilon_\mu \bm{\Psi}_\mu, \label{eq:Schrodinger}
\ee
we denote the value of the complex wave function $\bm{\Psi}_\mu$ at site $n$ by $\Psi_n$, and we observe that   
\be
\varepsilon_\mu \Psi_1 = -i\gamma\Psi_1 + t\Psi_2 
\ee
and  
\be
\varepsilon_\mu \Psi_2 =  i\gamma\Psi_2 + t(\Psi_1+\Psi_3) 
\ee
at the two sites on the right. They lead to
\be
\Psi_2 = \frac{\varepsilon_\mu+i\gamma}{t}\Psi_1 \label{eq:Psi2}
\ee
and 
\be
\Psi_3 = \left(-1+\frac{\varepsilon_\mu^2+\gamma^2}{t^2}\right)\Psi_1.\label{eq:Psi3}
\ee
By repeating the same calculations to the rest of the chain, we find 
\be
\Psi_1 + \Psi_5 = 2\alpha \Psi_3, 
\ee
and more generally,
\be
\Psi_n + \Psi_{n+4} = 2\alpha \Psi_{n+2}\label{eq:recur}
\ee
where 
\be
\alpha = \left(-1+\frac{{\varepsilon_\mu}^2+\gamma^2}{2t^2}\right)\label{eq:alpha}
\ee
and $n\in[1,2N-4]$ can be even or odd. Due to the symmetries of the finite-sized PT-dimer chain to be elucidated later, $\varepsilon_\mu$'s are again either real or imaginary, and hence $\alpha$ is always a real number. 

We note that Eq.~(\ref{eq:recur}) is a homogeneous linear recurrence relation, similar to that used to define the Fibonacci numbers. \cc{It was given in Ref.~\cite{linear} to show a linear localization phenomenon, where the amplitude of a non-Hermitian zero mode reduces linearly in space in a non-Hermitian reservoir made of coupled PT-dimers. The exact and explicitly quantized complex energies we derive below have not been reported previously.}  

We first focus on the odd-numbered sites and define $m=(n+1)/2$. Its solutions can be written in the form of 
\be
\Psi_n = e^{im\theta} + \beta_2 e^{-im\theta}\quad(n\; \text{odd}),\label{eq:general}
\ee
where $\theta\equiv \cos^{-1}\alpha$ is real (imaginary) if $|\alpha|<1$ ($|\alpha|>1$). By substituting Eq.~(\ref{eq:general}) with $m=1,2$ (i.e., for $\Psi_{1},\Psi_{3}$) into Eq.~(\ref{eq:Psi3}), we find
\be
\beta_2 = \frac{e^{2i\theta}[e^{i\theta}-(1+2\alpha)]}{e^{-i\theta}-(1+2\alpha)} = -e^{i\theta}.
\ee 
In the last step above, we used $\alpha=\cos\theta$ and the identity
\be
1+2\cos\theta = \frac{\sin\frac{3\theta}{2}}{\sin\frac{\theta}{2}}=\frac{e^{2i\theta}+e^{-i\theta}}{e^{i\theta}+1}.
\ee
Therefore, Eq.~(\ref{eq:general}) becomes
\be
\Psi_n = e^{im\theta} - e^{-i(m-1)\theta} \quad(n\; \text{odd}),\label{eq:general2}
\ee
and we find $\Psi_n\propto\sin(m\theta)$ in the large $m$ limit. 

To connect to the even-numbered lattice sites, we consider Eq.~(\ref{eq:Schrodinger}) on the leftmost site, which gives
\be
\Psi_{2N-1} = \frac{{\varepsilon_\mu}-i\gamma}{t}\Psi_{2N}
\ee  
similar to Eq.~(\ref{eq:Psi2}). Using $n=2N-1$ (and $m=N$) in Eq.~(\ref{eq:general2}) and we find $\Psi_{2N-1}=e^{iN\theta} - e^{-i(N-1)\theta}$, which gives
\be
\Psi_{2N} = t\frac{e^{iN\theta} - e^{-i(N-1)\theta}}{{\varepsilon_\mu}-i\gamma}.\label{eq:Psi2N}
\ee  
Similar to how we derived Eq.~(\ref{eq:general2}) on the odd-numbered lattice sites, we find the following expression for the odd-numbered lattice
\be
\Psi_n \propto e^{im'\theta} - e^{-i(m'-1)\theta} \quad(n\; \text{even}),\label{eq:general2b}
\ee
where $m'=(2N-n)/2+1$, i.e., the ordinal number of counting the even-numbered lattice sites from the left. Its normalization constant is fixed using Eq.~(\ref{eq:Psi2N}), i.e.,  
\begin{align}
\Psi_n &= t\frac{e^{iN\theta} - e^{-i(N-1)\theta}}{{\varepsilon_\mu}-i\gamma} \frac{e^{im'\theta} - e^{-i(m'-1)\theta}}{e^{i\theta} - 1} \quad(n\; \text{odd}) \nonumber \\
&=2it\frac{e^{i\theta/2}}{{\varepsilon_\mu}-i\gamma}\frac{ \sin\frac{2N-1}{2}\theta }{\sin\frac{\theta}{2}}\sin\left(\frac{2m'-1}{2}\theta\right) %&=i\frac{ \cos\left(\frac{2N}{2}-m\right)\theta - \cos\left(\frac{2N}{2}+m\right)\theta }{\sin\frac{\theta}{2}}\frac{te^{i\theta/2}}{{\varepsilon_\mu}-i\gamma}.
\label{eq:general3}
\end{align}

Finally, to obtain $\varepsilon_\mu$ analytically, we evaluate Eq.~(\ref{eq:general3}) with $n=2$ (and $m'=N$) and substitute it back into Eq.~(\ref{eq:Psi2}), which gives
\be
%\Psi_2=\frac{(e^{i2N\theta/2} - e^{-i(2N/2-1)\theta})^2}{e^{i\theta} - 1}\frac{t}{{\varepsilon_\mu}-i\gamma}=\frac{{\varepsilon_\mu}+i\gamma}{t}(e^{i\theta} - 1)
\Psi_2=2it\frac{e^{i\theta/2}}{{\varepsilon_\mu}-i\gamma}\frac{ \sin^2\frac{2N-1}{2}\theta }{\sin\frac{\theta}{2}}=\frac{{\varepsilon_\mu}+i\gamma}{t}(e^{i\theta} - 1)%=i\frac{{\varepsilon_\mu}+i\gamma}{t}2\sin\frac{\theta}{2}e^{i\theta/2}
\ee
or 
\be
\frac{\sin^2\frac{2N-1}{2}\theta}{\sin^2\frac{\theta}{2}} =\frac{{\varepsilon_\mu}^2+\gamma^2}{t^2}= 2(1+\cos\theta).\label{eq:sin2}
\ee 
It then leads to
\be
\sin^2\frac{2N-1}{2}\theta = (1+\cos\theta)(1-\cos\theta) = \sin^2\theta,
\ee 
and we find
\be
\frac{2N-1}{2}\theta = \pm\theta + q\pi\quad(q=1,2,\ldots,N).
\ee
The ``$+$'' sign turns out to be a false solution, and we arrive at
\be
\theta = \frac{2q\pi}{1+2N},\label{eq:theta}
\ee
or
\be
\varepsilon_\mu = \pm i \sqrt{\gamma^2-2t^2\left(1+\cos\frac{2q\pi}{1+2N}\right)}\label{eq:omega}
\ee
using the right half of Eq.~(\ref{eq:sin2}). By comparing with the band structure given by Eq.~(\ref{eq:E_periodic}), we identify $\theta$ as the discrete momentum $k$, and the energy quantization condition is just
\be
k(1+2N) = 2q\pi \quad(q=1,2,\ldots,N).
\ee
As we mentioned before, the spacing between two sites in a PT-dimer is taken to be 1 and dimensionless, and from the energy quantization condition above, we then find 
\be
L = 1+2N
\ee
to be the effective size of the concatenated PT-dimer chain. For a single PT-dimer, $N=1$ and hence its effective size is three instead of two in this regard. To verify that Eq.~(\ref{eq:omega}) still holds in this case, we observe that it reduces correctly to Eq.~(\ref{eq:E_dimer}) as anticipated, i.e.,
\be
\varepsilon_\mu = \pm i \sqrt{\gamma^2-2t^2\left(1+\cos\frac{2\pi}{3}\right)}=\pm i \sqrt{\gamma^2-t^2}.
\ee

\section{Validation}

\begin{figure*}
\includegraphics[clip,width=0.8\linewidth]{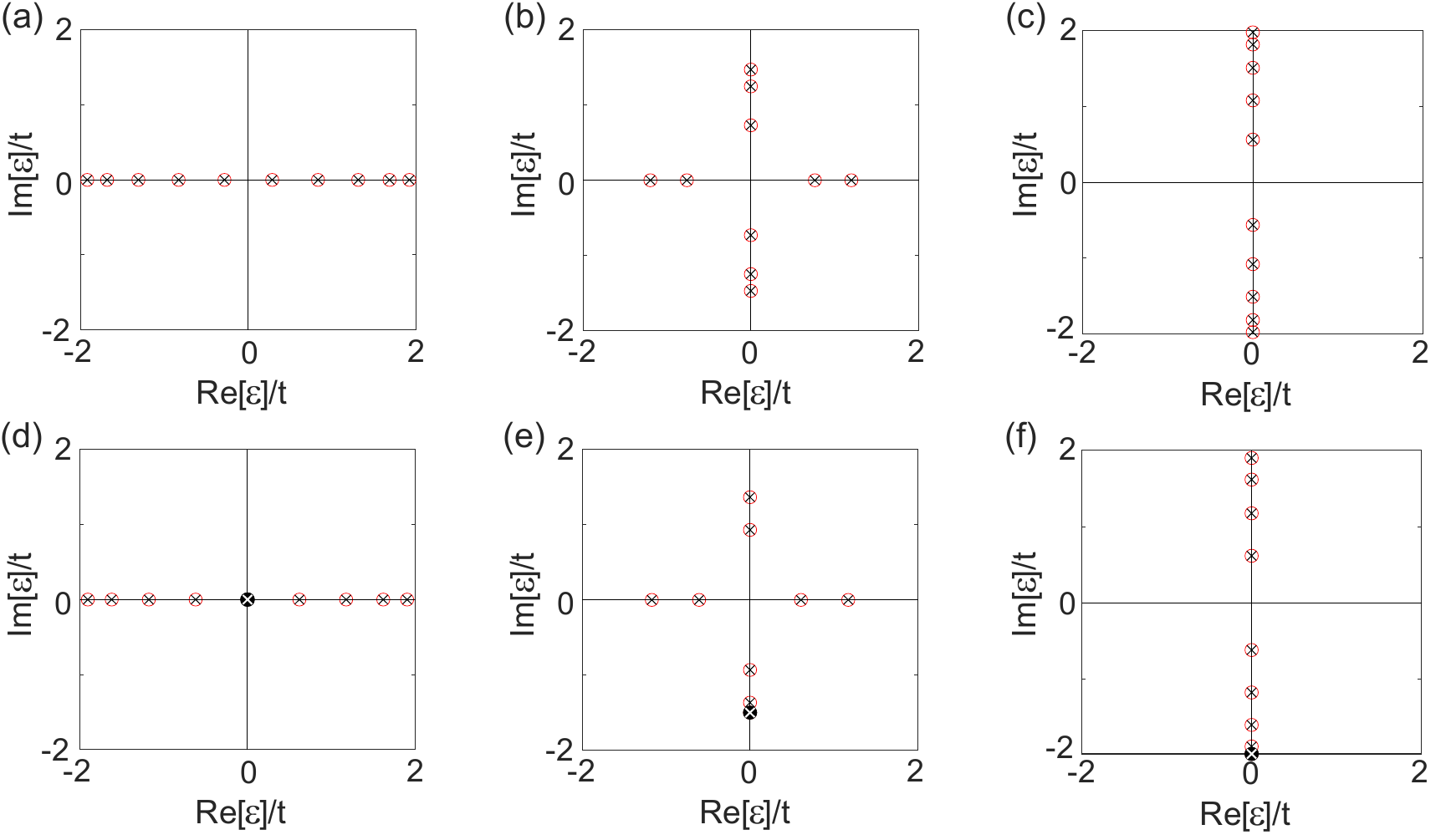}
\caption{(a-c) Complex energy levels of a concatenated PT-dimer chain with 5 PT-dimers. Open circles and crosses are numerical and analytical results, respectively. $\gamma/t=0$ in (a), 1.5 in (b), and 2 in (c). (d-f) Same as (a-c) but with the leftmost site removed. The unpaired modes are plotted using filled dots and white crosses.} \label{fig:N10}
\end{figure*}

Equation (\ref{eq:omega}) shows explicitly that $\varepsilon_\mu$'s are either real or imaginary, and to verify that this expression holds for all modes in the system, both in the PT-symmetric and PT-broken phases, we consider a chain with $N=5$ PT-dimers. At $\gamma=2t$, all modes are non-Hermitian zero modes \cite{zeromodeLaser}, with $\re{\varepsilon_\mu}=0$ and corresponding to the non-Hermitian flatband shown in Figs.~\ref{fig:E_periodic}(e,f). Equation~(\ref{eq:omega}) gives the correct $\varepsilon_\mu/t=\pm i0.563$, $\pm i1.081$, $\pm i1.512$, $\pm i1.819$, and $\pm i1.980$ (see Fig.~\ref{fig:N10}c). At $\gamma=1.5t$, equation~(\ref{eq:omega}) again gives the correct non-Hermitian zero modes at by $\varepsilon_\mu/t=\pm i0.731$, $\pm i1.249$, and $\pm i1.473$, as well as the real-valued ones at $\varepsilon_\mu/t=\pm0.762$ and $\pm1.197$ (see Fig.~\ref{fig:N10}b). Equation (\ref{eq:omega}) also works correctly in the Hermitian limit, as can be seen from Fig.~\ref{fig:N10}a.

Having validated the complex energy levels, we note a special property of $\theta$ (and $\alpha$) that we have not discussed so far. From Eq.~(\ref{eq:alpha}), it seems that $\alpha$ (and in turn $\theta$) is a function of $\gamma/t$ and $\mu$. However, as we have seen in Eq.~(\ref{eq:theta}), $\theta$ (and $\alpha$) is \textit{independent} of $\gamma/t$ and whether the corresponding $\varepsilon_\mu$ is real or imaginary. 

This observation points to a common misunderstanding of non-Hermitian lattices: a complex energy level does not indicate that the amplitude of the mode is amplified or attenuated \textit{across} the lattice. As Eqs.~(\ref{eq:general2}) and (\ref{eq:general2b}) show, the amplitude of an energy eigenstate grows or reduces exponentially across the lattice (i.e., as a function of $m$) only when $\theta$ has a finite imaginary part. Here however, $\theta$'s are all real as can be seen from Eq.~(\ref{eq:theta}). Therefore, the amplitudes of an energy eigenstate across the lattice only oscillate, at any given time or cross-section of coupled waveguides, between two bounds in a sinusoidal fashion, and these two bounds are given simply by $\pm1$ in Eq.~(\ref{eq:general2}) with the specified normalization. What is amplified or attenuated at a complex energy level, instead, are the amplitudes of an energy eigenstate in time or propagation inside the waveguides, where the propagation distance $z$ down the waveguides plays the role of time in the Schr\"odinger equation.

This distinction applies to other related wave phenomena in coupled waveguides as well. For example, if uniform straight waveguides are placed in parallel but with a random spacing, a propagation eigenstate of the coupled system can experience transverse Anderson localization \cite{transverseAnderson1,transverseAnderson2} and display localized mode profile across the lattice. However, as a propagation eigenstate, it does not exhibit any special spatial variation down the waveguides. Similarly, the linear localization across a non-Hermitian reservoir \cite{linear} mentioned previously does not display another linear amplitude change down the waveguides, because the linearly localized mode is not at an EP of the \cc{entire} system. 

\section{Multiple symmetries}

When introducing $\alpha$ in Eq.~(\ref{eq:alpha}), we mentioned that $\alpha$ is real using the property that $\varepsilon_\mu$'s must be real or imaginary as determined by the symmetries of the system. $\varepsilon_\mu$'s are real in the PT-symmetric phase, but PT symmetry does not warrant them to be imaginary in the PT-broken phase. This observation has not been appreciated in non-Hermitian photonics with quantum-inspired symmetries, and there is indeed another symmetry that pins the complex energy levels on the imaginary axis in the PT-broken phase.

This symmetry is termed non-Hermitian particle-hole symmetry \cite{zeromodeLaser} due to its identical consequence on the energy spectrum as that in a system with particle-hole symmetry in high-energy physics and condensed matter system. In its simplest form, a system with particle-hole symmetry satisfies the following anti-commutation relation 
\be
\{H,CK\} = CKH+HCK=0,\label{eq:NHPH}
\ee
where $C$ is a linear operator and $K$ is complex conjugation. It then indicates that if the (Hermitian) system has an eigenstate with energy $\varepsilon_\mu$, then it must have another eigenstate with energy 
\be
\varepsilon_\nu=-\varepsilon_\mu^*\label{eq:NHPH}
\ee
unless $\varepsilon_\mu=0$. Since the energy levels are real in a Hermitian system, the complex conjugation on $\varepsilon_\mu$ plays no role, but it does play a crucial role in a non-Hermitian system we consider here. Specifically, Eq.~(\ref{eq:NHPH}) indicates that $\varepsilon_\mu$ is on the imaginary axis if the two indices $\mu=\nu$.

For a PT-dimer, the symmetry operator $C$ can be chosen as the Pauli matrix 
\be
\sigma_z = 
\begin{pmatrix}
1 & 0 \\
0 & -1
\end{pmatrix}.
\ee
Its multiplication with the parity operator 
\be
P=\sigma_x = 
\begin{pmatrix}
0 & 1 \\
1 & 0
\end{pmatrix}
\ee
gives rise to another symmetry, i.e., non-Hermitian chiral symmetry \cite{chiral}:
\be
\{H,\Pi\} = \Pi H+H \Pi=0,\quad
\Pi = 
\begin{pmatrix}
0 & 1 \\
-1 & 0 \\
\end{pmatrix}
=i\sigma_y,
\label{eq:NHPH}
\ee
which implies that 
\be
\varepsilon_{\mu'}=-\varepsilon_\mu\label{eq:chiral}
\ee
or the complex energy spectrum is symmetric about the origin of the complex plane. These symmetries exist in the concatenated PT-dimer chain as well, with slightly more complicated symmetry operators. 

As we saw in Sec.~\ref{sec:quantization}, these symmetries do not actually play a role in our derivation of the energy quantization condition. Nevertheless, they do make our result easier to interpret as both $\alpha$ and $\theta$ are real. Interestingly, even if we lift PT symmetry (and consequently, also non-Hermitian chiral symmetry) explicitly by removing the leftmost site in Fig.~\ref{fig:schematic}b, the energy levels of our concatenated chain are still either real or imaginary, as we demonstrate below. 

In this case, we have $2N-1$ sites in the lattice, and the system is parity symmetric, with
\be
\Psi_{2N-1} = \pm\Psi_{1}.
\ee
As a result, the derivation of the energy quantization condition is much simpler by noting that $\Psi_{2N-1}$ and $\Psi_{1}$ are both described by Eq.~(\ref{eq:general2}), i.e., 
\be
e^{iN\theta} - e^{-i\left(N-1\right)\theta} = \pm (e^{i\theta} - 1)
\ee
or 
\be
\sin \left(N-\frac{1}{2}\right)\theta =\pm \sin\frac{\theta}{2}.
\ee
It leads to
\be
(2N-1)\theta = 2q\pi \pm \theta,
\ee
and again the ``$+$'' sign is a false solution. Therefore, we find that the $2N-1$ energy levels are given by
\be
\varepsilon_\mu = \pm i \sqrt{\gamma^2-2t^2\left(1+\cos\frac{q\pi}{N}\right)} \quad(q=1,2,\ldots,N-1),\nonumber
\ee
and
\be
\varepsilon_{2N-1} = - i \sqrt{\gamma^2-2t^2\left(1+\cos\frac{N\pi}{N}\right)} = -i\gamma. 
\ee
In other words, they are still either real or imaginary as we mentioned (see Figs.~\ref{fig:N10}(d-f)). Furthermore, they are still paired with just a sign difference except for $\varepsilon_{2N-1}$. 

We mention in passing that this difference from the PT-symmetric structure leads to a dramatic change of the eigenvalues of the scattering matrix \cite{CPALaser_PRA}, which connects the system to incoming and outgoing channels. More specifically, PT symmetry warrants that when a pole of the scattering matrix reaches the real frequency axis, it coincides with a zero of the same scattering matrix \cite{Longhi}. This property leads to a CPA-laser \cite{CPALaser} where CPA stands for coherent perfect absorption \cite{CPA,CPA_exp}. Interestingly, removing one loss layer from one end of a PT-symmetric optical heterostructure (corresponding to our removal of the leftmost site when $\gamma$ is negative) does not affect the operation of a CPA-laser \cite{CPALaser_exp}, but the linewidth displayed by the eigenvalues of the scattering matrix can change by orders of magnitude \cite{CPALaser_PRA}.

To find the effective size of the non-PT chain here with $2N-1$ sites, we use the momentum $k$ in place of $\theta$ as before, and we write the energy quantization condition as 
\be
k(2N) = 2q\pi \quad(q=1,2,\ldots,N),
\ee 
indicating that the system size is 
\be
L=2N \label{eq:L_defect}
\ee
despite that there are only $2N-1$ lattice sites.
\\
\vspace{10pt}

\section{Discussions}

In summary, we discussed the effective size of concatenated PT-dimer chains and the symmetry properties of their spectra, both with and without a boundary defect by removing a boundary site. In the limit that the number of PT-dimers goes to one, we find that the effective size of this single PT-dimer is three instead of two, despite the absence of apparent size-related factors in the energy levels. Similarly, when there is one and half PT-dimers, we find that its effective size is four instead of three from Eq.~(\ref{eq:L_defect}).  

To understand this difference of one between the effective size of the chain and the number of lattice sites, we revisit the quantization procedure in a 1D box potential in quantum mechanics. There we impose the Dirichlet boundary condition on the two boundaries, because it is assumed that the particle described by the quantum wave function cannot escape from the box, outside of which the potential is taken to be infinite. Here for a dimer and a finite-sized chain, apparently there lacks such a boundary condition, at least explicitly. Sometimes this situation is referred to as an open boundary condition, even though this term takes a different meaning in optics and photonics, where light can propagate from the lattice region to the external region. 

\begin{figure}[h]
\includegraphics[clip,width=\linewidth]{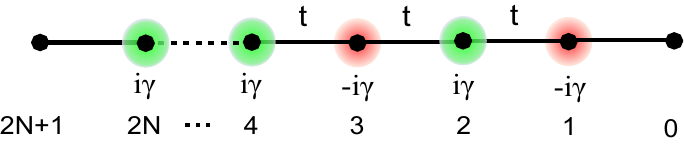}
\caption{Implicit Dirichlet boundary condition on two external lattice sites ``0'' and ``$2N+1$''.} \label{fig:boundary}
\end{figure}

However, there is actually a Dirichlet boundary condition implicit in our system. It is imposed outside the lattice region, at sites ``0'' and ``$2N+1$'' shown in Fig.~\ref{fig:boundary}. By requiring $\Psi_0 = \Psi_{2N+1}=0$, the Schr\"odinger equation for the lattice region, and in particular, at sites $1$ and $2N$, are the same as if these two additional lattice sites do not exist. Now the length between these two actual boundary points is $1+2N$, which is the effective size of the original lattice and one more than the number of sites in the latter. This knowledge is most likely known in some context for a Hermitian system, but it has not been explicitly and analytically checked in a non-Hermitian system, where the exact quantization condition is more difficult, if not impossible, to obtain due to the complex nature of its spectrum \cite{Bender2}. 

\acknowledgements

\cc{This work is supported by the National Science Foundation (NSF) under grant No. PHY-1847240.}

\begin{thebibliography}{99}

\bibitem{Bender1} C.~M.~Bender and S.~Boettcher,
%\textit{Real Spectra in Non-Hermitian Hamiltonians Having $\cal PT$ Symmetry},
Phys. Rev. Lett. {\bf 80}, 5243 (1998).
\bibitem{Bender2} C.~M.~Bender, S.~Boettcher, and P.~N.~Meisinger,
%\textit{$\cal PT$-Symmetric Quantum Mechanics},
J. Math. Phys. {\bf 40}, 2201 (1999).


\bibitem{Guo} A. Guo, G. J. Salamo, D. Duchesne, R. Morandotti, M. Volatier-Ravat, V. Aimez, G. A. Siviloglou, and D. N. Christodoulides, % Observation of PT-Symmetry Breaking in Complex Optical Potentials, 
Phys. Rev. Lett. \textbf{103}, 93902 (2009).
\bibitem{Hodaei} H. Hodaei, M.-A. Miri, M. Heinrich, D. N. Christodoulides, and M. Khajavikhan, %Parity-Time–Symmetric Microring Lasers, 
Science \textbf{346}, 975 (2014).
\bibitem{Peng} B. Peng, \c{Ş}. K. \"Ozdemir, F. Lei, F. Monifi, M. Gianfreda, G. L. Long, S. Fan, F. Nori, C. M. Bender, and L. Yang, %Parity-Time-Symmetric Whispering-Gallery Microcavities, 
Nat. Phys. \textbf{10}, 394 (2014).
\bibitem{Brandstetter} M. Brandstetter, M. Liertzer, C. Deutsch, P. Klang, J. Sch\"oberl, H. E. T\"ureci, G. Strasser, K. Unterrainer, and S. Rotter, %Reversing the Pump Dependence of a Laser at an Exceptional Point, 
Nat. Commun. \textbf{5}, 4034 (2014).


\bibitem{NPReview} L. Feng, R. El-Ganainy, and L. Ge, %``Non-Hermitian photonics based on parity-time symmetry,"
Nat. Photonics \textbf{11}, 752--762 (2017).

\bibitem{Bender_Exp} C. M. Bender, B. K. Berntson, D. Parker, and E. Samuel, %Observation of PT Phase Transition in a Simple Mechanical System, 
American J. Phys. \textbf{81}, 173 (2013).

\bibitem{Lawrence} M. Lawrence, N. Xu, X. Zhang, L. Cong, J. Han, W. Zhang, and S. Zhang, %Manifestation of PT Symmetry Breaking in Polarization Space with Terahertz Metasurfaces, 
Phys. Rev. Lett. \textbf{113}, 093901 (2014).

\bibitem{acoustic} R. Fleury, D. Sounas, and A. Al\'u, %An invisible acoustic sensor based on parity–time symmetry. 
Nat. Commun. \textbf{6}, 5905 (2015).

\bibitem{electronic} S. Assawaworrarit, X. Yu, and S. Fan, %Robust Wireless Power Transfer Using a Nonlinear Parity–Time-Symmetric Circuit, 
Nature \textbf{546}, 387 (2017).


\bibitem{EPreview} M.-A. Miri and A. Al\'u, %Exceptional points in optics and photonics,
Science \textbf{363}, 42 (2019).


\bibitem{reveal} H.-Z. Chen et al., %Revealing the Missing Dimension at an Exceptional Point, 
Nat. Phys. \textbf{16}, 571 (2020).

\bibitem{steering} H. Zhao, X. Qiao, T. Wu, B. Midya, S. Longhi, and L. Feng, %Non-Hermitian Topological Light Steering, 
Science \textbf{365}, 1163 (2019).

\bibitem{ptchain} K. Takata and M. Notomi, %Photonic Topological Insulating Phase Induced Solely by Gain and Loss, 
Phys. Rev. Lett. \textbf{121}, 213902 (2018).


\bibitem{Poli} C. Poli, M. Bellec, U. Kuhl, F. Mortessagne, and H. Schomerus, %Selective Enhancement of Topologically Induced Interface States in a Dielectric Resonator Chain, 
Nat. Commun. \textbf{6}, 6710 (2015).

\bibitem{Parto} M. Parto, S. Wittek, H. Hodaei, G. Harari, M. A. Bandres, J. Ren, M. C. Rechtsman, M. Segev, D. N. Christodoulides, and M. Khajavikhan, %Edge-Mode Lasing in 1D Topological Active Arrays, 
Phys. Rev. Lett. \textbf{120}, 113901 (2018).
\bibitem{Zhao} H. Zhao, P. Miao, M. H. Teimourpour, S. Malzard, R. El-Ganainy, H. Schomerus, and L. Feng, %Topological Hybrid Silicon Microlasers, 
Nat. Commun. \textbf{9}, 981 (2018).
\bibitem{Pan} [1] M. Pan, H. Zhao, P. Miao, S. Longhi, and L. Feng, %Photonic Zero Mode in a Non-Hermitian Photonic Lattice, 
Nat. Commun. \textbf{9}, 1308 (2018).

% \bibitem{St-Jean} P. St-Jean, V. Goblot, E. Galopin, A. Lemaître, T. Ozawa, L. Le Gratiet, I. Sagnes, J. Bloch, and A. Amo, %Lasing in Topological Edge States of a One-Dimensional Lattice, Nat. Photonics \textbf{11}, 651 (2017).
\bibitem{Gezai} G. L\'evai, A. Sinha, and P. Roy, %An Exactly Solvable 𝒫𝒯 Symmetric Potential from the Natanzon Class, 
J. Phys. A: Math. Gen. \textbf{36}, 7611 (2003).
\bibitem{Dey} S. Dey, A. Fring, and L. Gouba, %Milne Quantization for Non-Hermitian Systems, 
J. Phys. A: Math. Theor. \textbf{48}, 40FT01 (2015).


\bibitem{zeromodeLaser} L. Ge, %``Symmetry-protected zero-mode laser with a tunable spatial profile,"
Phys. Rev. A \textbf{95}, 023812 (2017).

\bibitem{Malzard} S. Malzard, C. Poli, and H. Schomerus, %Topologically protected defect states in open photonic systems with non-Hermitian charge-conjugation and parity-time symmetry,
Phys. Rev. Lett. \textbf{115}, 200402 (2015).

\bibitem{NHFlatband_PR} L. Ge, Photon. Res. \textbf{6}, A10–A17 (2018).

\bibitem{NHFlatband_PRL} B. Qi, L. Zhang and L. Ge, %``Defect states emerging from a non-Hermitian flat band of photonic zero modes,"
Phy. Rev. Lett. \textbf{120}, 093901 (2018).

\bibitem{noether} J. D. H. Rivero and L. Ge, %Pseudochirality: A Manifestation of Noether’s Theorem in Non-Hermitian Systems, 
Phys. Rev. Lett. \textbf{125}, 083902 (2020).

\bibitem{chiral} J. D. H. Rivero and L. Ge, %Chiral Symmetry in Non-Hermitian Systems: Product Rule and Clifford Algebra, 
Phys. Rev. B \textbf{103}, 014111 (2021).

\bibitem{Klaiman} S. Klaiman, U. G\"unther, N. Moiseyev, Phys. Rev. Lett. \textbf{101}, 080402 (2008).

\bibitem{linear} B. Qi and L. Ge, Adv. Phys. Res. \textbf{2}, 2300066 (2023).

\bibitem{makris} K. G. Makris, R. El-Ganainy, and D. N. Christodoulides, %Beam Dynamics in PT Symmetric Optical Lattices, 
Phys. Rev. Lett. \textbf{100}, 103904 (2008).

\bibitem{transverseAnderson1} H. De Raedt, A. Lagendijk, P. de Vries, %Transverse localization of light. 
Phys. Rev. Lett. \textbf{62}, 47-50 (1989)
\bibitem{transverseAnderson2} T. Schwartz, G. Bartal, S. Fishman, and M. Segev, 
%Transport and Anderson Localization in Disordered Two-Dimensional Photonic Lattices, 
Nature \textbf{446}, 52 (2007).

\bibitem{CPALaser_PRA} L. Ge and L. Feng, %Contrasting Eigenvalue and Singular-Value Spectra for Lasing and Antilasing in a PT -Symmetric Periodic Structure, 
Phys. Rev. A \textbf{95}, 013813 (2017).

\bibitem{Longhi} S. Longhi, %\textit{PT-Symmetric Laser Absorber}, 
Phys. Rev. A \textbf{82}, 031801(R) (2010).

\bibitem{CPALaser} Y.~D.~Chong, L.~Ge, and A.~D.~Stone, %\textit{$\cal PT$-Symmetry Breaking and Laser-Absorber Modes in Optical Scattering systems}, 
Phys. Rev. Lett. {\bf 106}, 093902 (2011).

\bibitem{CPA} Y. D. Chong, L. Ge, H. Cao, and A. D. Stone, %\textit{Coherent Perfect Absorbers: Time-Reversed Lasers}, 
Phys. Rev. Lett. \textbf{105}, 053901 (2010).
\bibitem{CPA_exp} W. Wan, Y. D. Chong, L. Ge, H. Noh, A. D. Stone, and H. Cao, %\textit{Time-Reversed Lasing and Interferometric Control of Absorption}, 
Science \textbf{331}, 889 (2011).


\bibitem{CPALaser_exp} Z. J. Wong, Y.-L. Xu, J. Kim, K. O’Brien, Y. Wang, L. Feng, X. Zhang, %\textit{Lasing and Anti-Lasing in a Single Cavity}, 
Nat. Photonics \textbf{10}, 796 (2016).




\end{thebibliography}
\end{document}